# A New Security and Identification Concept for SWIPT Systems in IoT Applications

Taki E. Djidjekh[#1], Lamoussa Sanogo[#2], Gaël Loubet[#3], Alassane Sidibé[#4], Daniela Dragomirescu[#5], Alexandru Takacs[#6]

[#]LAAS-CNRS, Université de Toulouse, CNRS, UPS, INSA

7, Avenue du Colonel Roche, 31400 Toulouse, France.

[1]tedjidjekh@laas.fr, [2]lsanogo@laas.fr, [3]loubet@laas.fr, [4]asidibe@laas.fr, [5]daniela@laas.fr, [6]atakacs@laas.fr

*Abstract*—This article addresses an innovative concept to enhance the security for IoT applications in the case of Simultaneous Wireless Information and Power Transfer. This is achieved by integrating a complementary security and identification mechanism through Wireless Power Transfer link within the network of autonomous wireless nodes. This mechanism is implemented at the level of the RF rectifier used to receive energy from a dedicated RF source. A prototype of such RF rectifier has been developed, it generates in real time a backscattered waveform (uplink from the wireless node to the RF source) as function of the parameters of the incoming energy waveform (downlink from RF source) and a digital private key code, generated/available at the level of the wireless node. This uplink waveform can be monitored at the level of the RF source for security/identification purposes implementing an autonomous hardware/physical security layer that operates independently from the communication protocols.

## I. INTRODUCTION

In today's landscape, the fast proliferation of Internet of Things (IoT) applications extends across diverse domains and integrates wireless sensor technologies in numerous sectors, such as Structural Health Monitoring (SHM), smart tracking, and smart factory. These systems rely on wireless connectivity, which makes them exceptionally vulnerable to cyber-attacks. These cyber-attacks are observed in all the IoT networks layers, including the wireless sensor nodes, by eavesdropping or replication [1]. Security mechanisms are already implemented in protocols such as LoRaWAN. Despite the complexity of their Medium Access Control (MAC) layers for authentication and cryptography, these protocols remain vulnerable to various cyber-attack scenarios [2]. Enhancing the security of these protocols requires increased computing resources at both node and gateway sides, leading to high energy consumption. Today, in the context of Simultaneous Wireless Information and Power Transfer (SWIPT) and Wireless Power Transfer (WPT) [3] the security mechanisms are implemented exclusively at the level of information transfer/communication while the energy link is used exclusively for WPT from RF source to the wireless nodes.

This article addresses an innovative and new concept to enhance the security of IoT applications by integrating a complementary security and identification mechanism through WPT link within the network of autonomous wireless nodes in a SWIPT context. To the author's best knowledges, this concept was never proposed before and open new research way in the field of emerging security concept for IoT applications.

## II. WPT BASED SECURITY AND IDENTIFICATION CONCEPT

The classical SWIPT approach is presented in Fig.1 with an emphasis on the electromagnetic waveform exchanged between the RF source and the wireless nodes in the context of a Wireless Sensors Network (WSN). Basically, a SWIPT WSN is composed of many wireless Sensors Nodes (SN), Communication Node(s) (CN) and RF source(s) (driven by the CN) interconnected in an ad-hoc mesh network. The Energy waveform (E-wave) generated by the RF source (DownLink (DL) to SN) has physical and digital characteristics and its unique feature is to provide wirelessly, at distance, power/energy to the SN. The proposed security and identification concept are implemented by: (i) adding an UpLink (UL) component (from SN to RF source) on E-wave; (ii) modifying in real time the digital content of this UL component of E-wave as function of a code provided by SN; (iii) adding an E-wave monitor on the level of RF source or CN.

In a SWIPT context the most effective solution to implement the beforementioned concept is to generate a backscattered waveform by modifying the topology of the RF rectifier of the SN. As illustrated in Fig 2, by introducing only one transistor (operating as an 'ON/OFF' switch) driven by the wireless System-on-Chip (SoC) (e.g. *via* UART or SPI interfaces) or by the Microcontroller Unit (MCU) of the SN, the matching between the rectifier and antenna can be dynamically degraded as function of a digital ON/OFF command signal (acting as a 'PriVate Key' (PVK)) applied to transistor gate. Thus, the instantaneous power/amplitude of the backscatter component is (in real time) dynamically modified as a function of digital (ON/OFF) PVK command provided by the SN. A table with a lot of PVK can be registered in the non-volatile memory of the SoC of the SN and CN. PVK can be randomly selected from this table or even generated by using a dedicated algorithm. E-wave monitor (topology and implementation not addressed in this paper) implemented on the CN/RF source will analyse, in a real-time manner, the UL (backscatter) component of the received E-wave by correlation with the DL component of the E-wave. DL component of E-wave can be regarded as a 'public key' implemented on physical/electromagnetic level. The Backscattering Rectifier (BR) proposed in Fig.2 is composed of: (i) an LC matching network with the Murata LQW18AN inductor (L = 43 nH) and C = 3.3 pF; (ii) a diode pair in a single package (SMS7630-005LF by Skyworks Solutions Inc); (iii) an N-channel MOSFET (BSS123 manufactured by ONSEMI); and (iv) an RC low pass filter (C2 = 100 pF and RL = 10 kΩ which simulates here the input impedance of the Power Management Unit (PMU) of the SN). The MOSFET operates in a diode-connected state, with a resistor R1 = 4 kΩ placed between the gate and drain. The source is connected to the anode of diode D2, while the drain-gate connection serves for control purposes. By adjusting the voltage applied to the MOSFET's gate, the drain-source ON-resistance changes, impacting the flow of current through diode D2 and the resistor R1. This manipulation effectively modifies the circuit's impedance, resulting in varying levels of reflection coefficient $S_{11}$ at the input port of the rectifier. It's essential to note that the circuit was initially optimized for matching when VGS CMD was set to 0 V. BR can operate in two modes: (i) energy harvesting mode: in this case BR charges, *via* the PMU, the storage element of the SN; and (ii) backscattering mode: in this case BR driven by a PVK provided by SN generates a digitally coded backscattered signal that is the UL component of the E-wave.

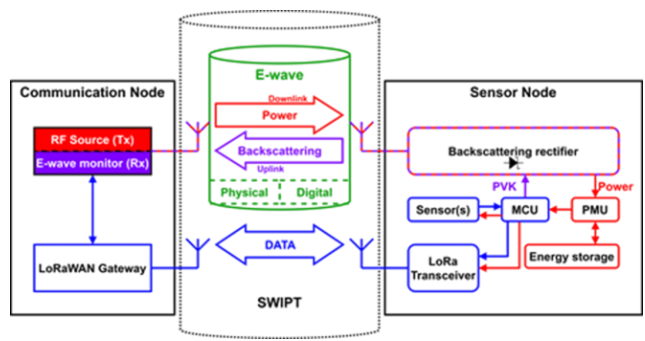

Fig.1. WPT based security and identification concept.

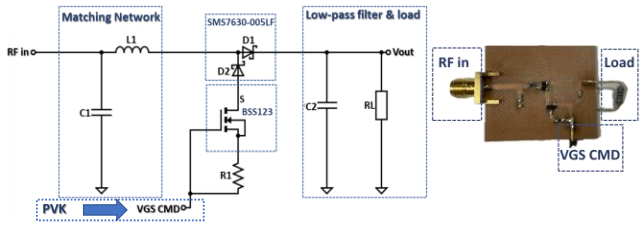

Fig.2. Backscattering rectifier circuit.

The 'classical topology' of the SN operating in a SWIPT context is also modified. As represented in Fig.1, the MCU commands by sending a PVK code to MOSFET's gate of the BR, which replaces the conventional rectifier topologies [3]. To facilitate reception of the downlink E-wave and transmission of the uplink/backscattered E-wave, two antennas (preferably operating with distinct polarizations) or a single antenna along with a circulator need to be employed. By leveraging this innovative approach, the SN's power consumption efficiency is practically the same. BR operates most of the time in a harvesting mode (harvesting power from the RF source *via* DL E-wave) and a very short time in the backscattering mode while transmitting the UL E-wave. This synchronized operation, piloted on the network level and facilitated by the MCU's commands to BR, marks a significant stride in enhancing the security wireless SN in SWIPT context.

III. EXPERIMENTAL RESULTS AND DISCUSSION

*1) Characterization of the Backscattering Rectifier in a Wired Setup*

The BR was designed to operate in ISM 868 MHz band, simulated in ADS (results not shown in this paper) and then manufactured in our lab by using a 2-layer PCB on FR4 substrate (thickness: 0.8 mm). First, the reflection coefficient at the RF input was measured using a Vector Network Analyzer (VNA) (Anristu MS4647A). The measurements were conducted at an RF power of -10 dBm when VGS CMD was set to 0 V (red trace Fig.3 (a)) and to 3.3 V (blue trace Fig.3 (a)).

Next, for setting up the harvesting characterization, we employed an RF signal generator (Anritsu MG3694B), connected directly to the rectifier's RF input. The voltage across the load RL = 10 kΩ (emulating the input impedance of the PMU) was measured using a high-precision multimeter (Keithley 2000). The experimental results were acquired using a home-made automated tool developed in LabVIEW.

Fig.3 (b) shows the DC voltage measured at the input port of the load (10 kΩ) as function of frequency (RF input power: -10 dBm) when VGS = 0 V (in this case the MOSFET is floating and consequently has a minimum impact on the rectifier behavior). The experimental results highlight that the best operating frequency of the BR is within the European 868 MHz ISM band and matches the ADS simulation results. To measure the rectifier's RF-to-DC power conversion efficiency, we set the continuous wave of the RF generator to the frequency of 868 MHz and sweep its power from -20 dBm to +20 dBm.

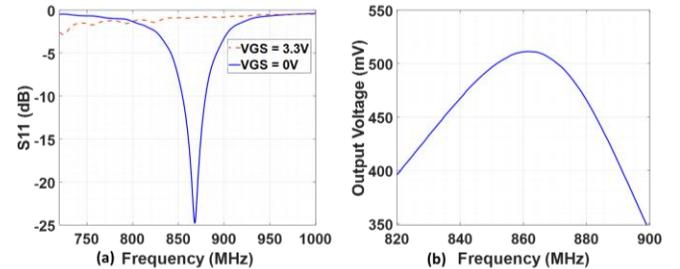

Fig.3. (a) Measurement of the rectifier reflection coefficient at -10 dBm RF input power (b) Output voltage measurement at the input port of a 10 kΩ load.

The obtained experimental results are represented in Fig.4.

The experiments to evaluating the backscattering behaviour were conducted by using a wired setup, represented in Fig.5, utilizing the RF signal generator (Anritsu MG3694B), an USB Spectrum Analyzer (Tektronix RSA306B), an RF circulator (Aerotek C11-1FFF) providing 20 dB as minimum isolation, and a waveform generator (Keithley 3390).

The RF input signal is connected to port 1 of the circulator, BR at port 2 and the Spectrum Analyzer at port 3 as illustrated in Fig.5 (b). The waveform generator was employed to drive the VGS CMD signal of the rectifier with a square signal

(frequency: 100 kHz, VLOW/OFF = 0 V and VHIGH/ON = 3.3 V). Thus, the backscattered signal of BR is ON/OFF modulated by VGS CMD signal. To prevent signal distortions the modulation/switching frequency was limited at around 100 kHz and the obtained experimental results are represented in Fig.6 for a RF input power of -15 dBm (CW at 876 MHz).

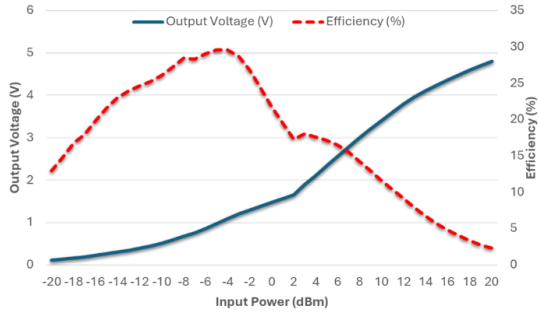

Fig.4. DC output voltage (blue line) and the RF-to-DC power conversion efficiency (red line) as function of RF input power (load: 10 kΩ).

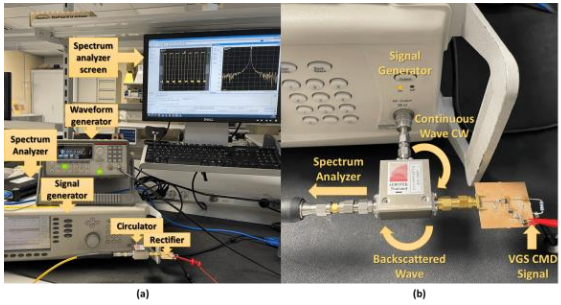

Fig.5: (a) Experimental setup; (b) Zoom on the circulator used to extract the backscattered signal.

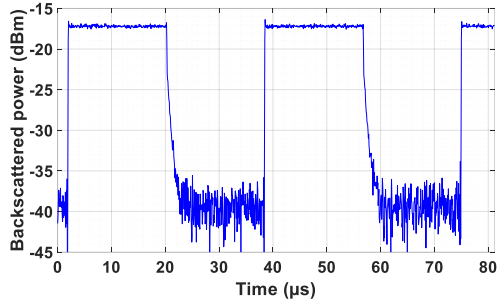

Fig.6. Backscattered signal measured at the BR port (wired setup) for an RF input power (continuous wave at 876 MHz) of -15 dBm and a modulating ON/OFF VGS CMD signal of 100 kHz.

The signal's dynamic range (difference between the high and low states) of the backscattered signal remains consistent as far as the switching frequency is lower than 100 kHz.

*2) Far-Field Wireless Measurements in an Anechoic Chamber by Using 3-Antennas Setup*

In this measurement stage, we utilized a three-antenna setup (represented in Fig.7) within an anechoic chamber in order to prevent any interferences with other experiments performed simultaneously in our lab. The rectifier was connected by using a patch antenna with a maximum gain of +9.2 dBi, positioned at 3.4 meters away from the WPT source, which employed a monopole antenna with a maximum gain of +2.5 dBi. The reception and analyse of backscattered signal were conducted by using the spectrum analyser connected to another patch antenna with a maximum gain of +9.2 dBi.

For minimizing the coupling between the WPT source and BR receiving antenna we aligned the BR receiving antenna with its radiation pattern's minimum towards the patch antenna, and insert an absorber wall between them. Consequently, this setup achieved a measured received signal level without the backscattered signal around -57 dBm for +15 dBm of RF power injected at the input of WPT antenna, which is lower than the backscattered signal. Then we conducted the same measurements as in stage 2 (wired setup) and the results of the received backscattered signal are shown in Fig.8 for different modulation frequency: (a) 10 kHz, (b) 100 kHz and (c) 1 kHz.

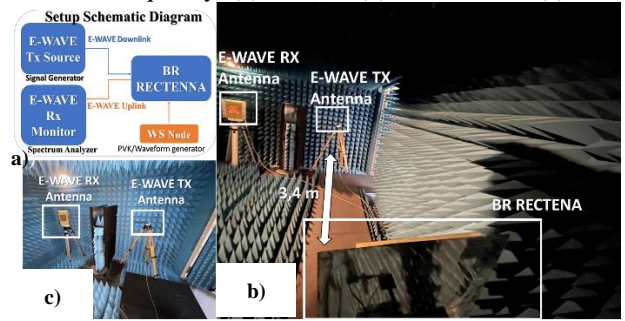

Fig.7. (a) Setup schematic diagram; (b) Photograph of the experimental setup: BR Rectenna (illuminated by E-wave Tx and controlled by a PVK/waveform generator), the Rx backscattered E-wave signal is monitored by the Spectrum Analyzer connected to E-wave Rx antenna; (c) Details on E-wave Tx and Rx antenna positioning.

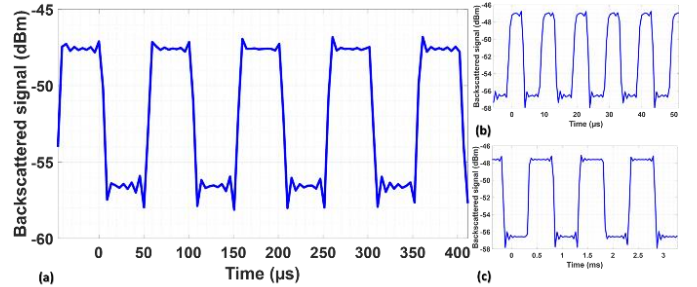

Fig.8. Backscattered signal detected by the spectrum analyser acting as E-wave monitor for a +15 dBm signal injected into the WPT antenna and: (a) a 10 kHz; (b) 100 kHz; and (c) 1 kHz; square signal injected at VGS CMD.

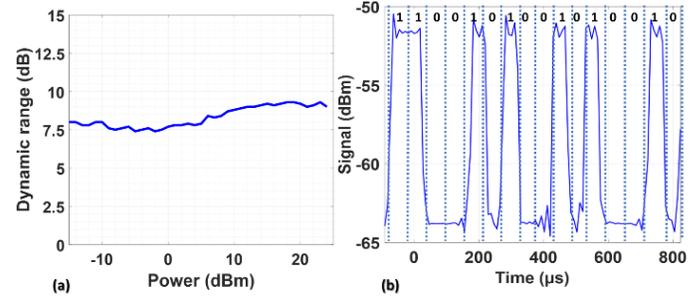

Fig.9. (a) Signal dynamic range as a function of RF power injected in E-wave Tx antenna; (b) Received E-wave Rx signal displayed on spectrum analyser for a PVK (random two-byte at 20 kHz) injected by a LoRa module on VGS CMD input of BR.

We further characterized the signal's dynamic range, varying the WPT power from -15 dBm to +24 dBm, as shown in Fig. 9 (a). The decrease in dynamic range did not exceed 1.5 dB. To verify the proper functioning of the BR in a wireless SN, we conducted tests using a programmable LoRaWAN module (MKR WAN 1310) to command the VGS CMD, by injecting a random two-bytes code at a frequency of 20 kHz, which was successfully retrieved, as depicted in Fig.9 (b).

*3) Discussion*

The experimental results for our BR prototype demonstrate a proof-of-concept for a simplified physical security layer model that seamlessly integrates into a wireless sensor node within the context of SWIPT. Backscattering techniques are intensively used mainly for RFID application, sensing and communication [5] but not for implementing a security mechanism at physical layer. A similar topology with the BR topology was proposed in [6] operating as harmonic transponder designed for sensing and identification applications. The integration of the proposed physical security layer occurs without extensive modifications of sensing node and without altering the communication standard (LoRaWAN in our case). The measurements illustrate the feasibility of transmitting and receiving a backscattered E-wave modulated (ON/OFF) with an PVK code and reaching distances of several meters with the respect of ISM 868 MHz EIRP regulation. Additionally, employing a 100 kHz modulation enables the transmission of multiple bytes of PVK within a few hundred microseconds, without significantly affecting energy consumption of the sensor node. Indeed, working in a real environment, outside of anechoic chamber, may degrade the dynamic range. Otherwise, employing two orthogonal polarizations (for UL and DL components of the E-wave) or two different harmonic frequency (using a harmonic transponder [6]-[7]) for transmission of the DL E-wave) and reception (UL E-wave) can significantly enhance the signal's dynamic range with the price of increasing system complexity.

## IV. Conclusion

This paper introduced a new and innovative concept to enhance the security and identification capabilities for IoT application in a SWIPT context. The proposed security mechanism exploits the waveform used for wireless power transfer and is complementary and independent from the security mechanism implemented by the communication protocols (e.g. LoRaWAN, BLE, etc.). The system architecture was described and the key hardware element that is the digitally controlled backscattered rectifier was designed and characterized. Experimental results have demonstrated that a digitally ON/OFF backscattered uplink energy waveform encrypting at electromagnetic level a local PVK can be generated by a wireless node, detected in real-time and in a far-field distance by an E-wave monitor for providing a complimentary security layer and legitimating the communication between wireless nodes.

## V. Acknowledgement

This work was partially supported by the French Research Agency (ANR) under grant agreement no. ANR-17-CE10-0014-03, the region OCCITANIE through the research project OPTENLOC and LAAS-CNRS micro and nanotechnologies platform, a member of the French Renatech network.